\documentclass[aps,prl,twocolumn]{revtex4}%
\usepackage{graphics}
\usepackage{amsmath}
\usepackage{graphicx}
\usepackage{amsfonts}
\usepackage{amssymb}
\usepackage{float}
\usepackage{longtable}
\usepackage{epsfig}
\usepackage{latexsym}
\usepackage{theorem}
\usepackage{bbm}
\usepackage{bm}
\usepackage{psfrag}%
\providecommand{\U}[1]{\protect\rule{.1in}{.1in}}
\begin{document}
\title{Performance of coherent-state quantum target detection\\in the context of asymmetric hypothesis testing}
\author{Gaetana Spedalieri and Stefano Pirandola}
\affiliation{Department of Computer Science, University of York, York YO10 5GH, UK}

\begin{abstract}
Due to the difficulties of implementing joint measurements, quantum
illumination schemes that are based on signal-idler entanglement are difficult
to implement in practice. For this reason, one may consider quantum-inspired
designs of quantum lidar/radar where the input sources are semiclassical
(coherent states) while retaining the quantum aspects of the detection. The
performance of these designs could be studied in the context of asymmetric
hypothesis testing by resorting to the quantum Stein's lemma. However, here we
discuss that, for typical finite-size regimes, the second- and third-order
expansions associated with this approach are not sufficient to prove quantum advantage.

\end{abstract}
\maketitle

%\affiliation{Computer Science and York Centre for Quantum
%Technologies, University of York, York YO10 5GH, UK}

%\begin{abstract}
%$M$-use $N$-ary quantum channel discrimination...
%\end{abstract}

\subsection{Introduction}

In coherent-state quantum target detection one exploits a semiclassical
source, specifically coherent states but a quantum detection scheme, not
necessarily homodyne or heterodyne detection (which are used
classically~\cite{radarBOOK}). This can therefore be considered a
quantum-inspired radar (QIR) since we relax the quantum properties of the
transmitter (i.e. no use of entanglement as in quantum
illumination~\cite{qill1,qill2,qill3,reviewSENSING,QCBGaussian}) while
retaining the optimal quantum performance of the receiver. We assume the
single-bin setting which corresponds to looking at some fixed range $R$ and
solving a binary test of target absent (null hypothesis $H_{0}$) or present
(alternative hypothesis $H_{1}$). We perform our study in the setting of
asymmetric hyphotesis testing~\cite{Stein1,Stein2,QHBound,GaeGAUSS}, so that
we fix the false-alarm probability to some reasonably low value, e.g.,
$p_{\text{FA}}=10^{-5}$, and then we minimize the probability of mis-detection
$p_{\text{MD}}$. Thus we look at the performance in terms of mis-detection
probability $p_{\text{MD}}$ versus signal-to-noise ratio (SNR) $\gamma$.

More precisely, these are the two quantum hypotheses to discriminate:

\begin{description}
\item[H$_{0}:$] A completely thermalizing channel, i.e., a channel with zero
transmissivity in an environment with $\bar{n}_{B}$\ mean thermal photons
(target absent).

\item[H$_{1}:$] A lossy channel with transmissivity $\eta$\ and thermal noise
$\bar{n}_{B}/(1-\eta)$, where the re-scaling avoids the possibility of a
passive signature (target present).
\end{description}

Let us consider an input coherent state $\left\vert \alpha\right\rangle $ with
mean number of photons $\bar{n}_{S}=|\alpha|^{2}$ and mean value
$\mathbf{\bar{x}}_{S}=(\bar{q},\bar{p})^{T}=\sqrt{2}(\operatorname{Re}%
\alpha,\operatorname{Im}\alpha)^{T}$. Without losing generality, we can assume
that $\alpha$ is real, so that $\mathbf{\bar{x}}_{S}=(\bar{q},\bar{p}%
)^{T}=\sqrt{2}(\alpha,0)^{T}$. On reflection from the potential target, we
have two possible output states:

\begin{description}
\item[H$_{0}:$] A thermal state $\rho_{0}^{\text{th}}$ with zero mean
$\mathbf{\bar{x}}_{0}=0$ and covariance matrix (CM) $\mathbf{V}_{0}=(\bar
{n}_{B}+1/2)\mathbf{I}$.

\item[H$_{1}:$] A displaced thermal state $\rho_{1}^{\text{th}}$ with mean
value $\mathbf{\bar{x}}_{1}=\sqrt{\eta}\mathbf{\bar{x}}_{S}$ and same
CM\ $\mathbf{V}_{1}=(\bar{n}_{B}+1/2)\mathbf{I}$.
\end{description}

Note that we have $\rho_{1}^{\text{th}}=D(\sqrt{\eta}\alpha)\rho
_{0}^{\text{th}}D(-\sqrt{\eta}\alpha)$ where $D$ is the phase-space
displacement operator.

\subsection{QIR performance}

In the setting of asymmetric hypothesis testing, the maximum performance
achievable by a QIR is given by the quantum Stein's lemma~\cite{Stein1,Stein2}%
. Suppose we want to discriminate between $M$ copies of two states, $\rho_{0}$
and $\rho_{1}$, using an optimal quantum measurement with output $k=0,1$. At
fixed false-alarm probability $p_{\text{FA}}:=p(1|\rho_{0}^{\otimes M})$, we
have the following decay of the false-negative (mis-detection) probability%
\begin{equation}
p_{\text{MD}}:=p(0|\rho_{1}^{\otimes M})\simeq\exp(-\beta M),
\label{firstORDER}%
\end{equation}
for some rate or error exponent $\beta$. According to the quantum Stein's
lemma, the optimal rate $\beta$ is equal to the relative entropy between the
two states, i.e.,
\begin{equation}
\beta=D\left(  \rho_{0}||\rho_{1}\right)  :=\mathrm{Tr}[\rho_{0}(\ln\rho
_{0}-\ln\rho_{1})].
\end{equation}

In a more refined version, we may account for second order asymptotics in $M$
and write~\cite{li2014second}
\begin{equation}
p_{\text{MD}}=e^{-MD\left(  \rho_{0}||\rho_{1}\right)  -\sqrt{MV\left(
\rho_{0}||\rho_{1}\right)  }\Phi^{-1}(p_{\text{FA}})+\mathcal{O}(\log M)},
\label{Li}%
\end{equation}
where we also use the quantum relative entropy variance%
\begin{equation}
V\left(  \rho_{0}||\rho_{1}\right)  =\mathrm{Tr}[\rho_{0}(\ln\rho_{0}-\ln
\rho_{1})^{2}]-[D\left(  \rho_{0}||\rho_{1}\right)  ]^{2},
\end{equation}
and the cumulative distribution function%
\begin{equation}
\Phi(\varepsilon):=\frac{1}{\sqrt{2\pi}}\int_{-\infty}^{\varepsilon}%
dx\exp\left(  -x^{2}/2\right)  ,
\end{equation}
with $\varepsilon\in(0,1)$ corresponding to (or bounding) the false-alarm
probability $p_{\text{FA}}$.

However, we need to notice that the term $\mathcal{O}(\log M)$ in
Eq.~(\ref{Li}) may play a non-trivial role in SNR calculations where $M$ is
not so large. According to Theorem~5 of Ref.~\cite{li2014second}, we have that
$\mathcal{O}(\log M)$ is between $0$ and $2\log M$, so that we have upper and
lower bounds for $p_{\text{MD}}$ (with quite some gap). A more refined
calculation involves to compute the third moment $T$ appearing in that
theorem. This will give more refined upper and lower bounds for the
performance of coherent states.

\subsection{First- and second-order terms}

We can write explicit formulas for the relative entropy $D\left(  \rho
_{0}||\rho_{1}\right)  $\ and the relative entropy variance $V\left(  \rho
_{0}||\rho_{1}\right)  $ of two arbitrary $N$-mode Gaussian states, $\rho
_{0}(\mathbf{\bar{x}}_{0},\mathbf{V}_{0})$ and $\rho_{1}(\mathbf{\bar{x}}%
_{1},\mathbf{V}_{1})$. The first one is given by~\cite{PLOB}
\begin{equation}
D\left(  \rho_{0}||\rho_{1}\right)  =-\Sigma\left(  \mathbf{V}_{0}%
,\mathbf{V}_{0}\right)  +\Sigma\left(  \mathbf{V}_{0},\mathbf{V}_{1}\right)  ,
\end{equation}
where we have defined the function
\begin{equation}
\Sigma\left(  \mathbf{V}_{0},\mathbf{V}_{1}\right)  =\frac{\ln\mathrm{det}%
\left(  \mathbf{V}_{1}+\frac{i\mathbf{\Omega}}{2}\right)  +\mathrm{Tr}\left(
\mathbf{V}_{0}\mathbf{G}_{1}\right)  +\delta^{T}\mathbf{G}_{1}\delta}{2},
\end{equation}
with $\delta=\mathbf{\bar{x}}_{0}-\mathbf{\bar{x}}_{1}$ and $\mathbf{G}%
_{1}=2i\boldsymbol{\Omega}\coth^{-1}\left(  2i\mathbf{V}_{1}\boldsymbol{\Omega
}\right)  $ being the Gibbs matrix~\cite{BanchiPRL}. The second one is given
by~\cite{Berta,RevQKD}%
\begin{equation}
V\left(  \rho_{0}||\rho_{1}\right)  =\frac{\mathrm{Tr}\left[  (\mathbf{\Gamma
}\mathbf{V}_{0})^{2}\right]  }{2}+\frac{\mathrm{Tr}\left[  (\mathbf{\Gamma
}\mathbf{\Omega})^{2}\right]  }{8}+\delta^{T}\mathbf{G}_{1}\mathbf{V}%
_{0}\mathbf{G}_{1}\delta,
\end{equation}
where $\mathbf{\Gamma}=\mathbf{G}_{0}-\mathbf{G}_{1}$. Using the output
states, $\rho_{0}^{\text{th}}$ and $\rho_{1}^{\text{th}}$, it is easy to
compute%
\begin{align}
D  &  :=D\left(  \rho_{0}^{\text{th}}||\rho_{1}^{\text{th}}\right)  =\eta
\bar{n}_{S}\ln(1+\bar{n}_{B}^{-1})\nonumber\\
&  =\gamma\bar{n}_{B}\ln(1+\bar{n}_{B}^{-1}),\\
V  &  :=V\left(  \rho_{0}^{\text{th}}||\rho_{1}^{\text{th}}\right)  =\eta
\bar{n}_{S}(2\bar{n}_{B}+1)\ln^{2}(1+\bar{n}_{B}^{-1})\nonumber\\
&  =\gamma\bar{n}_{B}(2\bar{n}_{B}+1)\ln^{2}(1+\bar{n}_{B}^{-1}),
\end{align}
where $\gamma:=\eta\bar{n}_{S}/\bar{n}_{B}$\ is the SNR. Note that, for large
background noise $\bar{n}_{B}\gg1$, we can expand%
\begin{equation}
D\simeq\gamma+\mathcal{O}(\bar{n}_{B}^{-1}),~V\simeq2\gamma+\mathcal{O}%
(\bar{n}_{B}^{-2}).
\end{equation}

Following Ref.~\cite[Theorem~5]{li2014second}, we may write the following
(approximate) bounds%
\begin{equation}
\frac{\Lambda}{M^{2}}\lesssim p_{\text{MD}}\lesssim\Lambda, \label{B1}%
\end{equation}
where
\begin{equation}
\Lambda:=\exp\left[  -MD-\sqrt{MV}\Phi^{-1}(p_{\text{FA}})\right]  .
\label{B2}%
\end{equation}
The upper bound in Eq.~(\ref{B1}) is the tool typically used in the
literature, while the lower bound is not taken into account (despite the gap
between the two bounds can become quite large).

\subsection{Computation of the third-order moment}

A more accurate version of Eq.~(\ref{B1}) includes higher-order terms and
suitable conditions of validity. Following Ref.~\cite{li2014second}, let us
introduce the third-order (absolute) moment%
\begin{equation}
T\left(  \rho_{0}||\rho_{1}\right)  =\sum_{x,y}\left\vert \left\langle
a_{x}\right\vert \left.  b_{y}\right\rangle \right\vert ^{2}\alpha
_{x}\left\vert \ln\frac{\alpha_{x}}{\beta_{y}}-D\left(  \rho_{0}||\rho
_{1}\right)  \right\vert ^{3}, \label{Tmoment}%
\end{equation}
where we use the spectral decompositions of the states
\begin{equation}
\rho_{0}=\sum_{x}\alpha_{x}\left\vert a_{x}\right\rangle \left\langle
a_{x}\right\vert ,~\rho_{1}=\sum_{y}\beta_{y}\left\vert b_{y}\right\rangle
\left\langle b_{y}\right\vert .
\end{equation}
See Appendix~\ref{AppT} for more details about the notation behind the formula
in Eq.~(\ref{Tmoment}).

Let $0<C<0.4748$ be the constant in the Berry--Esseen
theorem~\cite{Ccostant,Cconstant2}. Then, we may then write the more accurate
bounds~\cite[Theorem~5]{li2014second}%
\begin{gather}
\frac{1}{2^{9}M^{2}}\exp\left[  -MD\left(  \rho_{0}||\rho_{1}\right)
-\sqrt{MV\left(  \rho_{0}||\rho_{1}\right)  }\Phi^{-1}(\theta_{L})\right]
\nonumber\\
\leq p_{\text{MD}}\leq\label{bbb}\\
\exp\left[  -MD\left(  \rho_{0}||\rho_{1}\right)  -\sqrt{MV\left(  \rho
_{0}||\rho_{1}\right)  }\Phi^{-1}\left(  \theta_{U}\right)  \right]
,\nonumber
\end{gather}
where%
\begin{align}
\theta_{L}  &  :=p_{\text{FA}}+\frac{1}{\sqrt{M}}\left(  \frac{CT}{V\left(
\rho_{0}||\rho_{1}\right)  ^{3/2}}+2\right)  ,\\
\theta_{U}  &  :=p_{\text{FA}}-\frac{1}{\sqrt{M}}\frac{CT}{V\left(  \rho
_{0}||\rho_{1}\right)  ^{3/2}}.
\end{align}
More precisely, the bounds in Eq.~(\ref{bbb}) are valid as long as $M$ is
large enough to guarantee that $\theta_{L}\leq1$ and $\theta_{U}\geq0$, so
that they fall in the domain of $\Phi^{-1}$. From Eq.~(\ref{bbb}), we can
again notice how the lower bound become lose for increasing $M$.

Lat us compute the third moment $T$ for the output states $\rho_{0}%
^{\text{th}}$ and $\rho_{1}^{\text{th}}$, associated with the two hypothesis
(see Introduction). We have the following number-state spectral decompositions%
\begin{align}
\rho_{0}^{\text{th}}  &  =\sum_{k=0}^{\infty}\gamma_{k}\left\vert
k\right\rangle \left\langle k\right\vert ,~~\gamma_{k}:=\frac{\bar{n}_{B}^{k}%
}{(\bar{n}_{B}+1)^{k+1}},\\
\rho_{1}^{\text{th}}  &  =D(\sqrt{\eta}\alpha)\rho_{0}^{\text{th}}%
D(-\sqrt{\eta}\alpha)\nonumber\\
&  =\sum_{k=0}^{\infty}\gamma_{k}\left\vert k,\sqrt{\eta}\alpha\right\rangle
\left\langle k,\sqrt{\eta}\alpha\right\vert ,
\end{align}
whre $\left\vert k,\sqrt{\eta}\alpha\right\rangle =D(\sqrt{\eta}%
\alpha)\left\vert k\right\rangle $ is a displaced number state.

Using these decompositions in Eq.~(\ref{Tmoment}), we find
\begin{align}
&  T\left(  \rho_{0}^{\text{th}}||\rho_{1}^{\text{th}}\right)  =\sum
_{k,l=0}^{\infty}\left\vert \left\langle k\right\vert \left.  l,\sqrt{\eta
}\alpha\right\rangle \right\vert ^{2}\gamma_{k}\left\vert \ln\frac{\gamma_{k}%
}{\gamma_{l}}-D\left(  \rho_{0}||\rho_{1}\right)  \right\vert ^{3}\nonumber\\
&  =\sum_{k,l=0}^{\infty}\left\vert \left\langle k\right\vert D(\sqrt{\eta
}\alpha)\left\vert l\right\rangle \right\vert ^{2}\gamma_{k}\left\vert
\ln\frac{\gamma_{k}}{\gamma_{l}}-D\left(  \rho_{0}||\rho_{1}\right)
\right\vert ^{3}.
\end{align}
Because%
\begin{align}
D\left(  \rho_{0}^{\text{th}}||\rho_{1}^{\text{th}}\right)   &  =\eta\bar
{n}_{S}\ln\left(  \frac{\bar{n}_{B}+1}{\bar{n}_{B}}\right)  ,\\
\frac{\gamma_{k}}{\gamma_{l}}  &  =\frac{\bar{n}_{B}^{k-l}}{(\bar{n}%
_{B}+1)^{k-l}},\\
\ln\frac{\gamma_{k}}{\gamma_{l}}  &  =(k-l)\ln\left(  \frac{\bar{n}_{B}}%
{\bar{n}_{B}+1}\right)  ,
\end{align}
we may simplify%
\begin{align}
T\left(  \rho_{0}^{\text{th}}||\rho_{1}^{\text{th}}\right)   &  =\sum
_{k,l=0}^{\infty}\left\vert \left\langle k\right\vert D(\sqrt{\eta}%
\alpha)\left\vert l\right\rangle \right\vert ^{2}\\
&  \times\gamma_{k}\left\vert (k-l+\eta\bar{n}_{S})\ln\left(  \frac{\bar
{n}_{B}}{\bar{n}_{B}+1}\right)  \right\vert ^{3}.\nonumber
\end{align}

Now recall that~\cite[Eq.~(3.30) and Appendix~B]{Glauber}%
\begin{equation}
\left\langle k\right\vert D(\alpha)\left\vert l\right\rangle =\sqrt{\frac
{l!}{k!}}\alpha^{k-l}e^{-|\alpha|^{2}/2}\mathcal{L}_{l}^{(k-l)}(|\alpha|^{2}),
\end{equation}
where $\mathcal{L}_{n}^{(m)}(x)$ is an associated Laguerre polynomial, which
takes the following form in terms of the binomial coefficient~\cite{Szego}%
\begin{equation}
\mathcal{L}_{n}^{(m)}(x):=\sum_{k=0}^{n}\left(
\begin{array}
[c]{c}%
n+m\\
n-k
\end{array}
\right)  \frac{(-x)^{k}}{k!}.
\end{equation}
Therefore, for $\bar{n}_{S}=|\alpha|^{2}$, we may compute
\begin{equation}
\left\vert \left\langle k\right\vert D(\sqrt{\eta}\alpha)\left\vert
l\right\rangle \right\vert ^{2}=\frac{l!}{k!}(\eta\bar{n}_{S})^{k-l}%
e^{-\eta\bar{n}_{S}}\left[  \mathcal{L}_{l}^{(k-l)}(\eta\bar{n}_{S})\right]
^{2},
\end{equation}
so that we find the analytical expression%
\begin{align}
T\left(  \rho_{0}^{\text{th}}||\rho_{1}^{\text{th}}\right)   &  =e^{-\eta
\bar{n}_{S}}\sum_{k,l=0}^{\infty}\frac{l!}{k!}\gamma_{k}(\eta\bar{n}%
_{S})^{k-l}\left[  \mathcal{L}_{l}^{(k-l)}(\eta\bar{n}_{S})\right]
^{2}\nonumber\\
&  \times\left\vert (k-l+\eta\bar{n}_{S})\ln\left(  \frac{\bar{n}_{B}}{\bar
{n}_{B}+1}\right)  \right\vert ^{3}.
\end{align}
Note that this expression can be put in terms of the SNR $\gamma=\eta\bar
{n}_{S}/\bar{n}_{B}$ and the thermal background $\bar{n}_{B}$, i.e., we may
equivalently write%
\begin{align}
T\left(  \rho_{0}^{\text{th}}||\rho_{1}^{\text{th}}\right)   &  =e^{-\gamma
\bar{n}_{B}}\sum_{k,l=0}^{\infty}\frac{l!}{k!}\gamma_{k}(\gamma\bar{n}%
_{B})^{k-l}\left[  \mathcal{L}_{l}^{k-l}(\gamma\bar{n}_{B})\right]
^{2}\nonumber\\
&  \times\left\vert (k-l+\gamma\bar{n}_{B})\ln\left(  \frac{\bar{n}_{B}}%
{\bar{n}_{B}+1}\right)  \right\vert ^{3}. \label{bbbCOM}%
\end{align}
Furthermore, suitable bounds might be used for the Laguerre polynomials (see
Appendix~\ref{AppLAG}).

\subsection{Numerical investigation}

In order to perform a numerical comparison, we consider the error exponent%
\begin{equation}
\varepsilon_{\text{MD}}:=\frac{-\ln p_{\text{MD}}}{M},
\end{equation}
which corresponds to $\beta$ in Eq.~(\ref{firstORDER}) at the first order. It
is clear that the higher is the value of $\varepsilon_{\text{MD}}$, the better
is the discrimination performance.

\begin{figure}[t]
\vspace{0.2cm}
\par
\begin{center}
\includegraphics[width=0.40\textwidth] {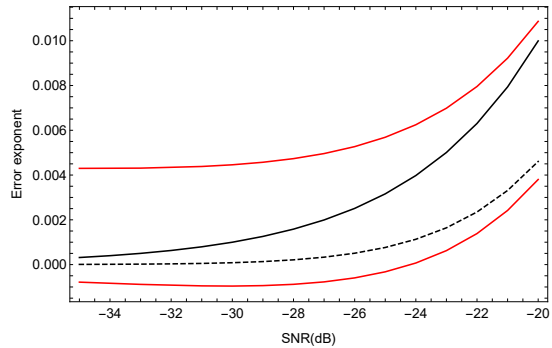}
\end{center}
\par
\vspace{-0.5cm}\caption{Error exponent $\varepsilon_{\text{MD}}$ as a function
of the SNR in dBs $10\log_{10}\gamma$. We compare the first order
approximation of Eq.~(\ref{firstORDER}) (black line), with the higher-order
lower and upper bounds from Eq.~(\ref{bbb}) (red lines). We also plot the
Marcum benchmark (dashed black line). We consider the parameters
$p_{\text{FA}}=10^{-3}$, $M=5000$ and $\bar{n}_{B}=600$. Note that the lower
bound even becomes negative for lower values of SNRs.}%
\label{scanPIC}%
\end{figure}

To show the finite-size behavior, we consider $p_{\text{FA}}=10^{-3}$,
$M=5000$ and bright background\ $\bar{n}_{B}=600$. With these parameters, we
plot $\varepsilon_{\text{MD}}$\ versus SNR in decibels (i.e., $10\log
_{10}\gamma$) for the optimized detection for coherent states considering the
first order formula of Eq.~(\ref{firstORDER}), and the higher-order bounds in
Eq.~(\ref{bbb}). As a comparison, we also plot the error exponent achievable
by a classical radar which employs coherent state pulses and heterodyne
detection~\cite{radarBOOK}. This can be computed from the Marcum
Q-function~\cite{Marcum,albersheim1981closed}
\begin{align}
p_{\text{MD}}  &  =1-Q\left(  \sqrt{2\gamma},\sqrt{-2\ln p_{\text{FA}}%
}\right)  ,\\
Q(x,y)  &  :=\int_{y}^{\infty}dt~te^{-(t^{2}+x^{2})/2}I_{0}(tx),
\end{align}
with $I_{0}(.)$ being the modified Bessel function of the first kind of zero order.

As we can see from Fig.~\ref{scanPIC}, the QIR would have a clear advantage
over the Marcum benchmark if we consider the asymptotic first order formula.
However, the first order expression of Eq.~(\ref{firstORDER}) is valid only
for very large $M$. For a typical finite size value of $M$, we need to
consider the higher-order bounds in Eq.~(\ref{bbb}), but we see that the gap
is too large to reach a conclusion of quantum advantage.

\subsection{Conclusion}

In this work, we have studied a quantum-inspired lidar/radar based on coherent
states and optimal quantum detection, analysing the performance in the context
of asymmetric hypothesis testing (quantum Stein's lemma, higher-order
asymptotics). According to our study, the current mathematical tools do not
allow us to prove quantum advantage over classical strategies based on
coherent states and heterodyne detection when a finite number of probes is
considered. Such an advantage may be claimed in the asymptotic limit of very
large number of probes, so that the first order order becomes completely
dominant over the higher-order terms. However, such an asymptotic regime is
not relevant for practical applications.

\bigskip

\textbf{Acknowledgments}.~~This work has been funded by the European Union's
Horizon 2020 Research and Innovation Action under grant agreement No. 862644
(FETOpen project: Quantum readout techniques and technologies, QUARTET). S. P.
would like to thank Quntao Zhuang for discussions.

\appendix

\section{Relative entropy notation~\cite{QCBSkola}\label{AppT}}

Relative entropy is given by
\begin{equation}
D\left(  \rho_{0}||\rho_{1}\right)  :=\mathrm{Tr}[\rho_{0}(\ln\rho_{0}-\ln
\rho_{1})].
\end{equation}
Using the spectral decompositions
\begin{equation}
\rho_{0}=\sum_{x}\alpha_{x}\left\vert a_{x}\right\rangle \left\langle
a_{x}\right\vert ,~\rho_{1}=\sum_{y}\beta_{y}\left\vert b_{y}\right\rangle
\left\langle b_{y}\right\vert ,
\end{equation}
and therefore%
\begin{align}
\ln\rho_{0}  &  =\sum_{x}\ln\alpha_{x}\left\vert a_{x}\right\rangle
\left\langle a_{x}\right\vert ,\\
\ln\rho_{1}  &  =\sum_{y}\ln\beta_{y}\left\vert b_{y}\right\rangle
\left\langle b_{y}\right\vert ,
\end{align}
we may write
\begin{align}
D\left(  \rho_{0}||\rho_{1}\right)   &  =\sum_{x}\alpha_{x}\left\langle
a_{x}\right\vert (\ln\rho_{0}-\ln\rho_{1})\left\vert a_{x}\right\rangle \\
&  =\sum_{x}\alpha_{x}\left[  \ln\alpha_{x}-\sum_{y}\ln\beta_{y}\left\vert
\left\langle a_{x}\right\vert \left.  b_{y}\right\rangle \right\vert
^{2}\right]  .
\end{align}
Let us set $\left\vert a_{x}\right\rangle =\sum_{y}\gamma_{xy}\left\vert
b_{y}\right\rangle $ with complex $\gamma_{xy}$ such that $\sum_{x}\left\vert
\gamma_{xy}\right\vert ^{2}=\sum_{y}\left\vert \gamma_{xy}\right\vert ^{2}=1$.
Therefore,%
\begin{align}
D\left(  \rho_{0}||\rho_{1}\right)   &  =\sum_{x}\alpha_{x}\left(  \ln
\alpha_{x}-\sum_{y}\ln\beta_{y}\left\vert \gamma_{xy}\right\vert ^{2}\right)
\nonumber\\
&  =\sum_{x,y}\alpha_{x}\left\vert \gamma_{xy}\right\vert ^{2}\left(
\ln\alpha_{x}-\ln\beta_{y}\right) \nonumber\\
&  =\sum_{x,y}p_{xy}\ln\frac{\alpha_{x}}{\beta_{y}}:=\left\langle \ln
\frac{\alpha(X)}{\beta(Y)}\right\rangle ,
\end{align}
where $\alpha(X):=\{\alpha_{x},p_{x}\}$, $\beta(Y):=\{\beta_{y},p_{y}\}$ where
$p_{x}$ and $p_{y}$ are the marginal distributions of the joint probability
$p_{xy}:=\alpha_{x}\left\vert \gamma_{xy}\right\vert ^{2}$ which is the
probability to get $X=x$ and $Y=y$ by measuring $\rho_{0}$ in the basis
$\{\left\vert a_{x}\right\rangle \}$ and then in $\{\left\vert b_{y}%
\right\rangle \}$. In this notation, we may also write the relative entropy
variance as follows%
\begin{equation}
V\left(  \rho_{0}||\rho_{1}\right)  =\left\langle \ln\frac{\alpha(X)}%
{\beta(Y)}\right\rangle ^{2}-D\left(  \rho_{0}||\rho_{1}\right)  ^{2}.
\end{equation}
The third-order moment entering the quantum Stein's lemma is given
by~\cite{li2014second}
\begin{align}
T\left(  \rho_{0}||\rho_{1}\right)   &  =\left\langle \left\vert \ln
\frac{\alpha(X)}{\beta(Y)}-D\left(  \rho_{0}||\rho_{1}\right)  \right\vert
^{3}\right\rangle \\
&  =\sum_{x,y}\left\vert \left\langle a_{x}\right\vert \left.  b_{y}%
\right\rangle \right\vert ^{2}\alpha_{x}\left\vert \ln\frac{\alpha_{x}}%
{\beta_{y}}-D\left(  \rho_{0}||\rho_{1}\right)  \right\vert ^{3}.
\end{align}

\section{Useful bounds\label{AppLAG}}

Various bounds are known for the associated Laguerre polynomials. A well-known
uniform bound is the Szeg\"{o} bound~\cite{Szego}
\begin{equation}
\left\vert \mathcal{L}_{n}^{(m)}(x)\right\vert \leq\frac{(m+1)_{n}}{n!}%
e^{x/2},
\end{equation}
for $x,m\geq0,~n=0,1,\ldots$where we use the Pochhammer's symbol (or shifted
factorial)%
\begin{align}
(a)_{0}  &  =1,\\
(a)_{n}  &  =a(a+1)(a+2)\cdots(a+n-1),\\
(a)_{n}  &  =\frac{\Gamma(a+n)}{\Gamma(a)}%
\end{align}
with $\Gamma(a)$ being the Gamma function. Another one is~\cite{Rooney}
\begin{equation}
\left\vert \mathcal{L}_{n}^{(m)}(x)\right\vert \leq2^{-m}q_{n}e^{x/2},
\end{equation}
for $x\geq0$, $m\leq-1/2$, $n=0,1,\ldots$ and where we set%
\begin{equation}
q_{n}=\frac{\sqrt{(2n)!}}{2^{n+1/2}n!}\simeq\frac{1}{\sqrt[4]{4\pi n}%
}~\text{for large }n\text{.}%
\end{equation}

\end{document}